\documentclass[showpacs,nofootinbib,preprintnumbers,prd,superscriptaddress,twocolumn]{revtex4}

\usepackage{amsmath}
\usepackage{amsfonts}
\usepackage{amssymb}
\usepackage{graphicx}
\usepackage[titletoc]{appendix}
\usepackage{color}
\usepackage{hyperref}
\usepackage{cleveref}
\usepackage[rightcaption]{sidecap}
\usepackage{subfigure}
\usepackage{comment}

\usepackage{mathtools}

\usepackage{dcolumn}

\usepackage{array}
\usepackage{ctable}
\usepackage{multirow}
\usepackage{siunitx}
\usepackage{longtable}
\usepackage{tabularx}
\usepackage{booktabs}

\graphicspath{{Graphics/}}

\def\be{\begin{equation}}
\def\ee{\end{equation}}
\def\bea{\begin{eqnarray}}
\def\eea{\end{eqnarray}}

\definecolor{vividviolet}{rgb}{0.62, 0.0, 1.0}
\definecolor{amaranth}{rgb}{0.9, 0.17, 0.31}
\definecolor{palatinateblue}{rgb}{0.15, 0.23, 0.89}
\definecolor{brightpink}{rgb}{1.0, 0.0, 0.5}
\definecolor{cornflowerblue}{rgb}{0.39, 0.58, 0.93}
\definecolor{deepcarminepink}{rgb}{0.94, 0.19, 0.22}
\definecolor{radicalred}{rgb}{1.0, 0.21, 0.37}
\hypersetup{ linktoc=all,
    colorlinks, linkcolor={palatinateblue},
    citecolor={brightpink}, urlcolor={amaranth}
}

\begin{document}

\title{Entanglement area law violation from field-curvature coupling}

\author{Alessio Belfiglio}
\email{alessio.belfiglio@unicam.it}
\affiliation{School of Science and Technology, University of Camerino, Via Madonna delle Carceri, Camerino, 62032, Italy.}
\affiliation{Istituto Nazionale di Fisica Nucleare (INFN), Sezione di Perugia, Perugia, 06123, Italy.}

\author{Orlando Luongo}
\email{orlando.luongo@unicam.it}
\affiliation{School of Science and Technology, University of Camerino, Via Madonna delle Carceri, Camerino, 62032, Italy.}
\affiliation{Istituto Nazionale di Fisica Nucleare (INFN), Sezione di Perugia, Perugia, 06123, Italy.}
\affiliation{SUNY Polytechnic Institute, 13502 Utica, New York, USA.}
\affiliation{INAF - Osservatorio Astronomico di Brera, Milano, Italy.}
\affiliation{Al-Farabi Kazakh National University, Al-Farabi av. 71, 050040 Almaty, Kazakhstan.}

\author{Stefano Mancini}
\email{stefano.mancini@unicam.it}
\affiliation{School of Science and Technology, University of Camerino, Via Madonna delle Carceri, Camerino, 62032, Italy.}
\affiliation{Istituto Nazionale di Fisica Nucleare (INFN), Sezione di Perugia, Perugia, 06123, Italy.}

\begin{abstract}
We investigate the entanglement entropy of a massive scalar field nonminimally coupled to spacetime curvature, assuming a static, spherically symmetric background. We discretize the field Hamiltonian by introducing a lattice of spherical shells and imposing a cutoff in the radial direction. We then study the ground state of the field and quantify deviations from area law due to nonminimal coupling, focusing in particular on Schwarzschild-de Sitter and Hayward spacetimes, also discussing de Sitter spacetime as a limiting case. We show that large positive coupling constants can significantly alter the entropy scaling with respect to the boundary area, even in case of small field mass. Our outcomes are interpreted in view of black hole entropy production and early universe scenarios.
\end{abstract}

\pacs{03.65.Ud, 04.60.-m, 04.62.+v}

\maketitle


\section{Introduction} \label{intro}

In recent years, the importance of entanglement entropy in characterizing quantum systems has been widely recognized. Its significance extends to various fields, including quantum information processing \cite{ben1, ben2, ben3}, condensed matter physics \cite{condmat1, condmat2, condmat3, condmat4, condmat5, condmat6}, and quantum field theory \cite{qft1, qft2, qft3}. The amount of entanglement in a quantum state is also crucial in determining the complexity of numerical simulations involving that state\footnote{For an introduction to the complexity of quantum many-body systems and their simulation, see \cite{rev1} and references therein.}. It is commonly believed that the entropy of a distinguished region should always exhibit extensive scaling, as in the case of thermal states \cite{rev1}. However, for typical ground states of quantum systems, a different behavior emerges. These ground states typically follow an ``area law", where the entropy scales linearly with the boundary area of the region, accompanied by small corrections, often logarithmic in nature. This characteristic is intimately connected to the locality of interactions in quantum many-body systems \cite{rev1, loc1, loc2, loc3, loc4, loc5, loc6}.

To quantify this fact, the von Neumann entropy is often employed as a suitable figure of merit. It is widely used and serves as an operationally defined measure of entanglement for pure states \cite{qit, hor, pl1}. This entropy can be computed for various partitions of the system, providing insights into the quantum correlations among the chosen subparts. In many physical systems, the local degrees of freedom are arranged in specific geometric patterns, such as spin chains, networks, or D-dimensional lattices \cite{rev1, rev2}. These systems may also possess a continuum limit described by a quantum field theory, where the number of degrees of freedom is formally unbounded. 

To avoid divergences in such scenarios, a short-distance regulator is usually introduced, serving as ultraviolet cutoff. In Ref. \cite{qft1}, it was initially demonstrated that the entanglement entropy of a free scalar field in its ground state is proportional to $\mathcal{A}_H/a^2$, where $\mathcal{A}_H$ represents the boundary area of the region under consideration (the ``horizon"), and $a$ is the ultraviolet cutoff, i.e., the lattice spacing chosen in the model\footnote{Throughout the text, we employ natural units $\hbar=k_B=c=1$, thus measuring lengths in GeV$^{-1}$.}.  Subsequently, the same result was obtained by tracing over the degrees of freedom within a spherical surface \cite{qft2}. In both cases, the calculations were performed in a flat background. However, from a quantum field theory perspective, the interest in entanglement entropy primarily stems from potential connections with Bekenstein-Hawking black hole entropy \cite{bh1, bh2, bh3, bh4} and the holographic principle \cite{holo}. Hence, it is essential to generalize these results to curved spacetimes. 

Recently, it was demonstrated \cite{das} that an area law also applies to static, spherically symmetric configurations when the field is minimally coupled to the curvature of spacetime. However, the effects of nonminimal field-curvature coupling appear to be relevant in several relativistic contexts, e.g. early universe dynamics or regions near black holes \cite{faki, mae}. They also predict novel phenomena, such as particle creation \cite{blm1,blm2}, non equivalence between frames \cite{fara}, large-scale dynamical effects \cite{lm1}, vacuum energy cancellation \cite{lm2,lm3}, and so on. 
 
Accordingly, we here try to investigate area law in case of nonminimally coupled fields. We explore nonminimal coupling based on a Yukawa-like structure, where a scalar field is coupled to spacetime curvature. Specifically, to characterize our analysis, we focus on spacetime configurations that transport vacuum energy, distinguishing between singular and regular solutions. We emphasize such geometric structures for their great importance in cosmological contexts such as inflation and/or black hole physics. Indeed, we consider the widely-known Schwarzschild-de Sitter \cite{SDS} and Hayward \cite{hay} metrics, also discussing de Sitter spacetime as limiting case for both scenarios. Through our investigation, we demonstrate that nonminimal coupling can have a notable impact on the dependence of entropy on the area, suggesting that local interactions between quantum fields and the background geometry can play an important role in entanglement generation\footnote{Possible back-reaction mechanisms \cite{back1,back2,back3}, which can affect quantum field dynamics in some cosmological contexts, are neglected for the moment.}. To quantitatively assess these modifications, we numerically compute the entropy by discretizing the Hamiltonian derived from our field scenario, modelling it as a system of harmonic oscillators on a lattice of spherical shells. Such procedure allows us to critically compare our results with previous findings, highlighting the main deviations from area law due to field-curvature coupling. We will see that these effects are more relevant in case of large coupling constants or, alternatively, for large vacuum energy contributions, which are typical of early universe scenarios. 

The paper is structured as follows. In Sect. \ref{Sec1}, we review the standard computation of entanglement entropy for a system of coupled harmonic oscillators in its ground state. In Sect. \ref{Sec2}, we derive the scalar field Hamiltonian on a lattice of spherical shells, assuming a spherically symmetric background with field-curvature coupling. We show under which conditions the scalar field degrees of freedom can be modeled as harmonic oscillators on a flat background.  In Sect. \ref{sez4}, the area law is re-obtained in case of minimal coupling and deviations due to nonminimal coupling are discussed for Schwarzschild-de Sitter and Hayward spacetimes. In the same section, we explore physical consequences of our findings. Finally, in Sect. \ref{sez5}, we draw our conclusions and discuss future perspectives.


\section{Entanglement entropy of coupled harmonic oscillators}  \label{Sec1}

Our procedure requires to explore discrete field theories, instead of continuous systems. Discretized quantum systems in fact allow for uncontroversial computation of the entropy \cite{rev2}. This is not the case of quantum field theories, where renormalization and regularization are usually needed in order to avoid divergences. A common strategy is then to study quantum fields on a lattice, where the lattice spacing allows for ultraviolet regularization of the theory. 

A scalar field on a lattice can be modeled as an open chain of $N$ coupled harmonic oscillators, so for convenience we briefly review the derivation of the ground state entanglement entropy in this latter case. The Hamiltonian of the system reads 
\be \label{hoham}
H= \frac{1}{2} \sum_{i=1}^N p_i^2+ \frac{1}{2} \sum_{i,j=1}^N x_i K_{ij} x_j,
\ee
where the matrix $K$ is symmetric and has positive eigenvalues. The normalized ground-state wave function is then \cite{qft2}
\be \label{gswf}
\psi_0 (x_1, \dots, x_N)= \pi^{-N/4} (\text{det}\  \Omega)^{1/4} \exp{[-x \cdot \Omega \cdot x/2]},
\ee
with $\Omega := \sqrt{K}$. We trace now over the first $n$ harmonic oscillators of the system. For convenience, we write $\Omega$ in block form as
\be \label{oblk}
\Omega= \begin{pmatrix} A & B \\[4 pt]
B^T & C  \end{pmatrix},
\ee
where $A$ is a $n \times n$ matrix and $C$ is $(N-n) \times (N-n)$. 
Further defining 
\begin{subequations}
    \begin{align}
& \beta:= \frac{1}{2} B^T A^{-1} B  \label{bemat}\\  
& \gamma:= C- \beta \label{gamat},
    \end{align}
\end{subequations}
we can write the reduced density operator of the system in the position representation as \cite{qft2}
\be \label{redop}
\rho_{\rm out} (x, x^\prime) \simeq \exp{[-(x \cdot \gamma \cdot x+ x^\prime \cdot \gamma \cdot x^\prime)/2+ x \cdot \beta \cdot x^\prime]},
\ee
where now both $x$ and $x^\prime$ are vectors with $N-n$ components. Following consolidate approaches \cite{qft2, inver}, the von Neumann entropy for $\rho_{\rm out}$ reads 
\begin{align} \label{enten}
\mathcal{S}(\rho_{\rm out})&\equiv -{\rm Tr}\left( \rho_{\rm out} \ln \rho_{\rm out} \right)  \notag \\
&=\sum_{j=n+1}^{N} \left( -\ln (1-\nu_j) - \frac{\nu_j}{1-\nu_j} \ln \nu_j\right),
\end{align}
with 
\be \label{csiv}
\nu_i= \frac{\lambda_i}{1+\sqrt{1-\lambda_i^2}}
\ee
and $\lambda_i$ are the eigenvalues of the matrix $\gamma^{-1}\beta$.

The above treatment, despite general, is not developed in the context of curved spacetime. However, the effects of geometry can be significant in various cosmological scenarios, leading to departures from flat spacetime results.
Hence, we will now derive the discrete scalar field Hamiltonian in a simple curved spacetime, showing under which conditions it can be traced back to Eq. \eqref{hoham}. We focus in particular on static, spherically symmetric backgrounds, taking also into account possible field-curvature couplings.


\section{Scalar field theory on a lattice of spherical shells} \label{Sec2}

We start from the action of a massive scalar field $\varphi$ propagating in a background spacetime described by the metric tensor $g_{\mu \nu}$,
\be \label{action}
S= - \frac{1}{2} \int d^4x \sqrt{-g} \left[ g^{\mu \nu} \partial_\mu \varphi \partial_\nu \varphi + \left( m^2+\xi R \right) \varphi^2   \right],
\ee
where $m$ is the field mass, $R$ the Ricci scalar curvature and $\xi$ the coupling constant to curvature. The coupling here considered is a Yukawa-like interaction, developed by the two fields $\varphi$ and $R$ and prompting relevant results in both gravitational and cosmological contexts, see e.g. \cite{faki, mae, blm1}. 

The simplest approach consists in working out a static, spherically symmetric spacetime, described by the line element\footnote{We adopt the same conventions found in \cite{das, inver}, where the metric tensor is taken with positive signature.}
\be \label{linel}
ds^2= - f(r) dt^2+ \frac{dr^2}{g(r)}+ r^2(d\theta^2+\sin^2 \theta d\phi^2),
\ee
where in general $f(r)\neq g(r)$. A suitable example of such a choice is the Tolman-Oppeneimer-Volkov spacetime, widely-used to characterize compact objects, interiors or accretors \cite{accr1, accr2, accr3}.

The spacetime in Eq. \eqref{linel} can be further simplified in a Schwarzschild-like configuration, namely handling $f(r)=g(r)$, developing classes of solutions with zero or non-zero energy momentum tensors, that are widely investigated in the literature \cite{exact}. Exploiting spherical symmetry, we can decompose the scalar field $\varphi$ in terms of real spherical harmonics, to write 
\be \label{spharm}
\varphi(x)=\sum_{lm} \varphi_{lm}(t,r) Z_{lm}(\theta, \phi).
\ee
A suitable strategy is now to employ the Lema\^itre coordinates \cite{lila} 
\begin{subequations}
    \begin{align}
&\zeta= t + \int dr \frac{\left[ 1-f(r) \right]^{-1/2}}{f(r)} \label{lemr} \\
&\chi= t \pm \int dr \frac{\sqrt{1-f(r)}}{f(r)}  \label{lemt}
    \end{align}
\end{subequations}
and to fix Lema\^itre time $\chi =0$, so that the discrete scalar field Hamiltonian reads \cite{das}
\be \label{hamsum}
H= \sum_{lm} H_{lm}(0)
\ee
with
\begin{align} \label{hamexp}
H_{lm}(0)= \frac{1}{2} \int_0^\infty dr \bigg[& \frac{\pi_{lm} r^{-2}}{1-f(r)} + r^2 \left( \partial_r \varphi_{lm}\right)^2 \notag \\
&+ l(l+1)\varphi_{lm}^2+ r^2\left(m^2+\xi R\right) \varphi_{lm}^2\bigg].
\end{align}
The canonical momenta $\pi_{lm}$ satisfy the Poisson brackets
\be \label{pois}
\{ \varphi_{lm}(r), \pi_{lm}(r^\prime) \}= \sqrt{1-f(r)} \delta(r-r^\prime)
\ee
and, by means of the canonical transformation
\be \label{cantrasf}
\pi_{lm} \rightarrow r \sqrt{1-f(r)} \pi_{lm},\ \ \ \ \varphi_{lm} \rightarrow \frac{\varphi_{lm}}{r},
\ee
we arrive at
\begin{align} \label{radham}
H_{lm}(0)= \frac{1}{2} \int_0^\infty dr \bigg \{& \pi_{lm}^2(r)+ r^2 \bigg[ \frac{\partial}{\partial r}\left( \frac{\varphi_{lm}}{r} \right) \bigg]^2 \notag \\
&+ \left( \frac{l(l+1)}{r^2}+m^2+ \xi R \right) \varphi_{lm}^2 \bigg \}.
\end{align}
Remarkably, in the above equation we have the presence of the curavture $R$. With $f(r)=g(r)$, Eq. \eqref{linel} yields
\be \label{ricscal}
R(r)=\frac{2-2f(r)-4rf^\prime(r)-r^2 f^{\prime \prime}(r)}{r^2}.
\ee
The final step in our discretization procedure will consist in introducing a short-distance regulator, to provide an ultraviolet regularization of the radial coordinate $r$. In this way we will be able to describe the scalar field degrees of freedom in terms of the coupled harmonic oscillators studied in Sec. \ref{Sec1}.
\begin{figure*}
 \centering
\includegraphics[width=1.5\columnwidth,clip]{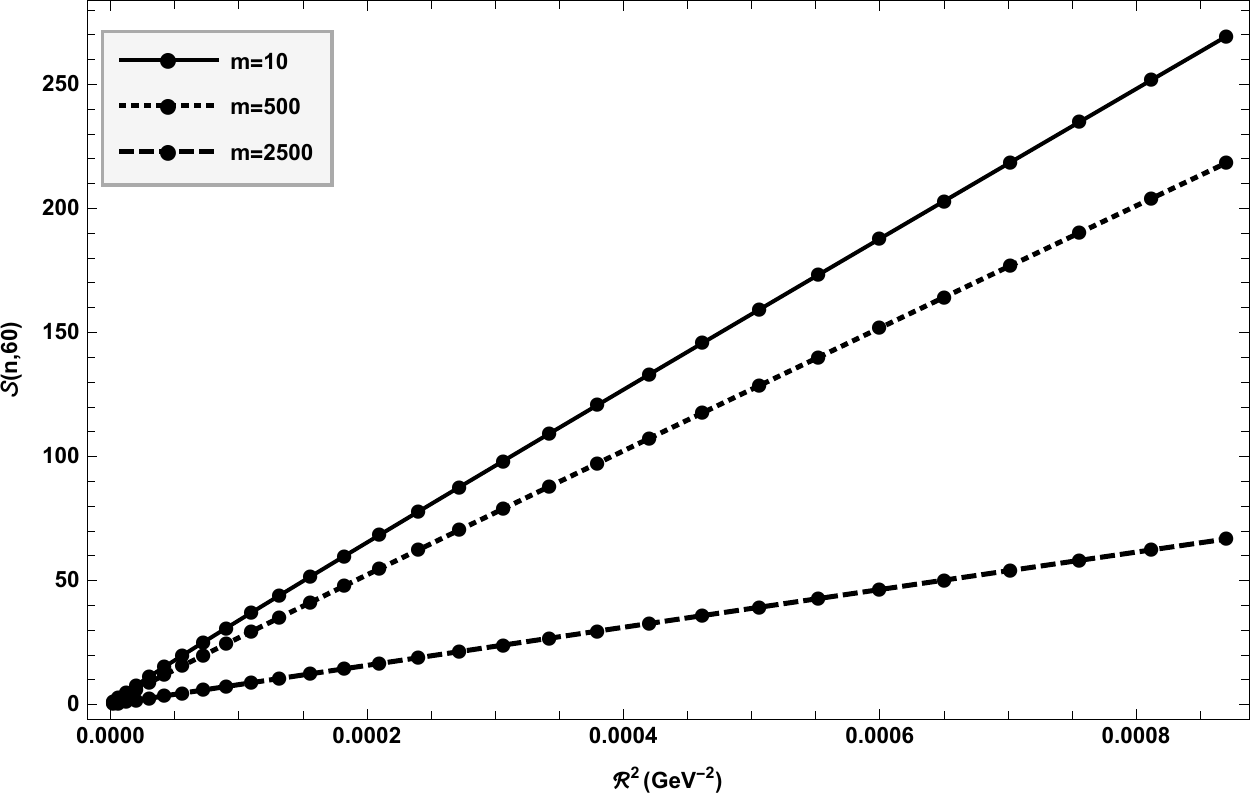}\hfill
    \caption{Entanglement entropy $\mathcal{S}(n,N)$ as function of $\mathcal{R}^2$ for $\xi=0$. The other parameters are: $a=10^{-3}$ GeV$^{-1}$, $l_{\rm max}=1000$, $N=60$ and $n < N/2$. We notice that $\mathcal{S} \propto R^2$, implying that an area law is satisfied in the minimal coupling scenario. The proportionality coefficient between entropy and area decreases for larger field masses.}
    \label{fnocurv}
\end{figure*}


\subsection{Ultraviolet regularization of the Hamiltonian} \label{Sec2A}
 
The only continuous variable left in Eq. \eqref{radham} is the radial coordinate $r$. At this point, we adopt the standard regularization scheme \cite{qft2, rev2}, which consists in introducing a lattice of $N$ spherical shells, spaced by a given distance, say $a$. Specifically, the latter represents one of the free parameters in the model, needful to increase (or decrease) the accuracy of our discretization procedure\footnote{Additional details and physical consequences on this approach can be also found in Ref. \cite{inver}.}.

The fully discrete Hamiltonian finally reads
\begin{align} \label{discham}
H= &\frac{1}{2a} \sum_{lm} \sum_{j=1}^N \bigg[ \pi_{lm,j}^2 + \left( j+\frac{1}{2} \right)^2  \left( \frac{\varphi_{lm,j+1}}{j+1}-\frac{\varphi_{lm,j}}{j} \right)^2  \notag \\
&+ \left( \frac{l(l+1)}{j^2}+ (m^2+\xi R)a^2  \right) \varphi_{lm,j}^2 \bigg],
\end{align}
where $\varphi_{lm,j} \equiv \varphi_{lm}(r_j)$ and $\pi_{lm,j} \equiv \pi_{lm} (r_j)$, with $r_j \equiv ja$.
Moreover, we need to discretize the scalar curvature in Eq. \eqref{ricscal}, hence we realize the following changes:
\begin{subequations}
    \begin{align}
& f^\prime(r) \rightarrow \frac{f_{j+1}-f_j}{a} \label{fpri} \\
& f^{\prime \prime}(r)\rightarrow \frac{f_{j+1}-2f_{j}+f_{j-1}}{a^2} ,  \label{fsec}
    \end{align}
\end{subequations}
where $f_j \equiv f(r_j)$. 

The total size of the system is now $A=(N+1)a$, where $a$ and $A$ act as ultraviolet and infrared cutoff, respectively. Hence, both parameters behave as cut off scales and therefore also fix the overall energy in our model.

\begin{figure*}[ht!]
 \centering
\includegraphics[scale=0.64]{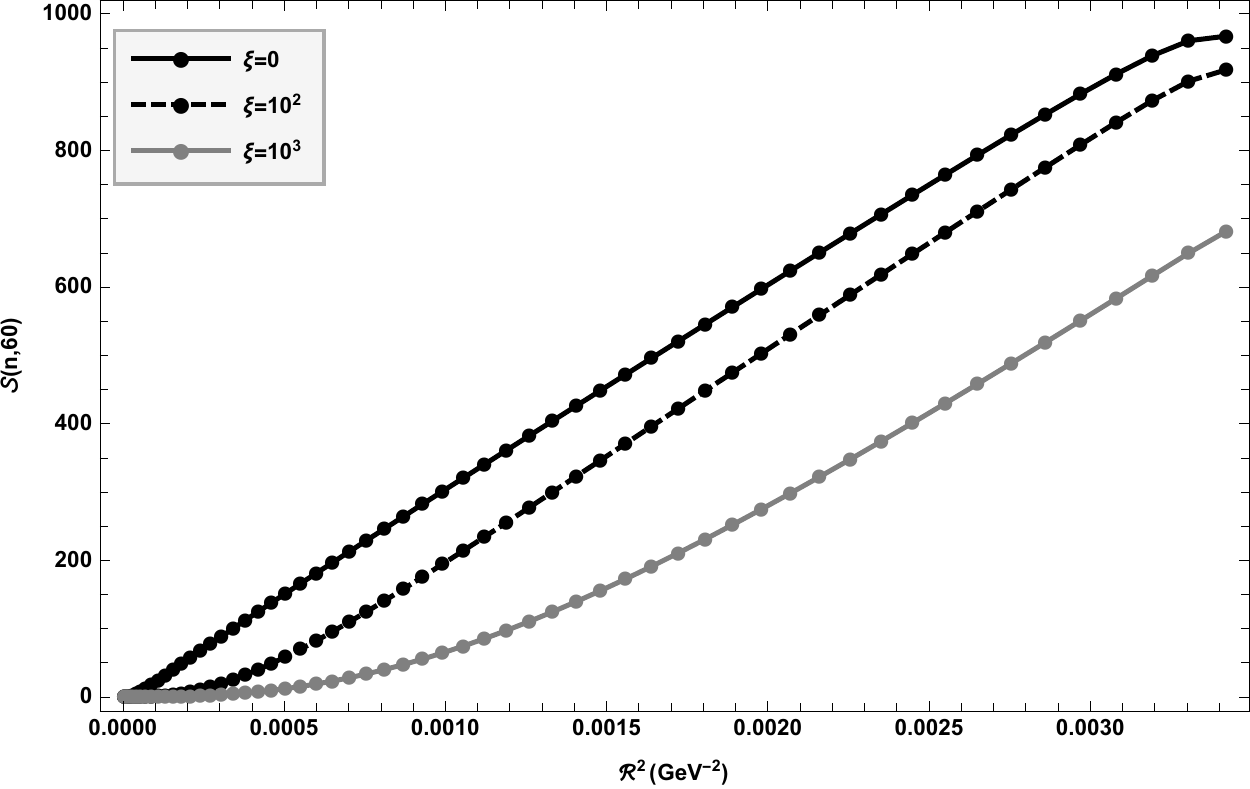}
\hspace{1cm}
\includegraphics[scale=0.64]{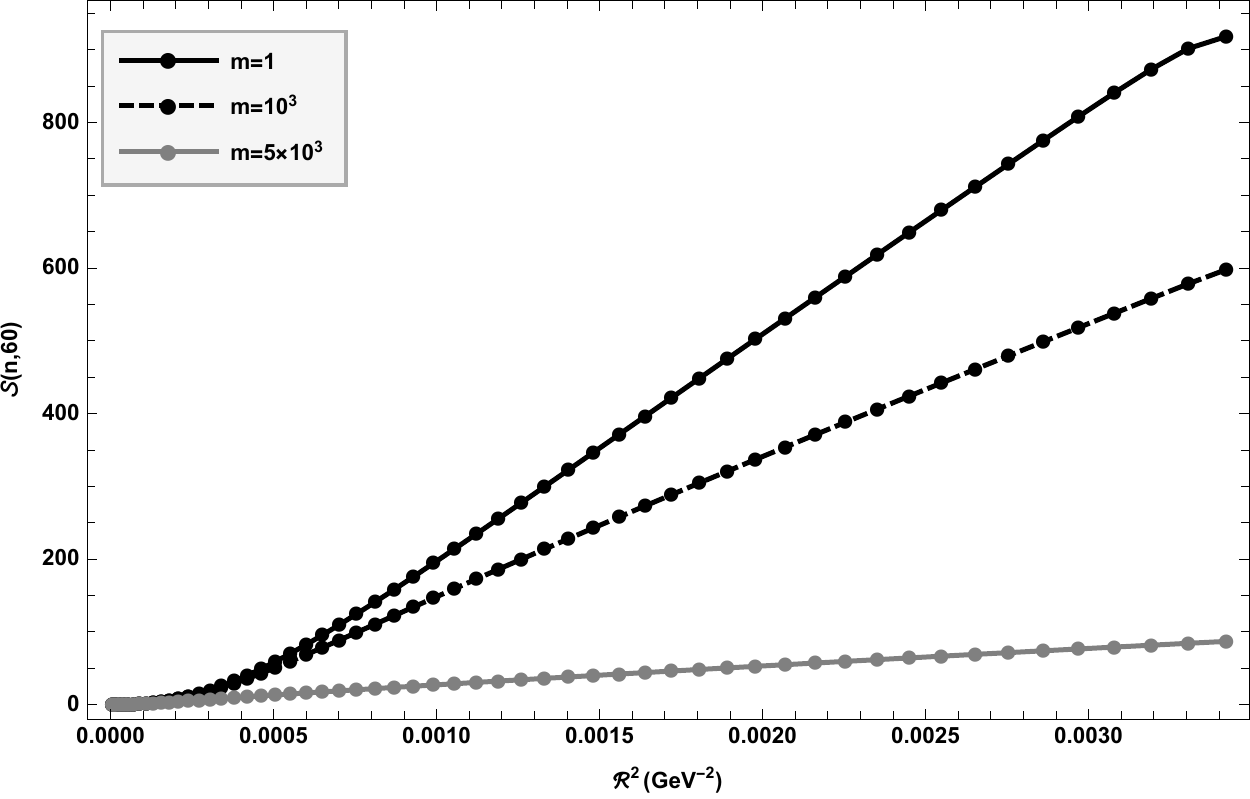}
        \caption{Entanglement entropy $\mathcal{S}(n,N)$ in Schwarzschild-de Sitter spacetime for different values of the field-curvature coupling constant $\xi$ (on the left, with $m=1$ GeV) and field mass $m$ (on the right, with $\xi=10^2$). The other parameters are: $a=10^{-3}$ GeV$^{-1}$, $l_{\rm max}=1000$, $N=60$, $n< N-1$, $GM=1$ GeV$^{-1}$ and $L=1000$ GeV$^{-1}$. As $n \rightarrow 1$, the entanglement entropy does not scale linearly in $\mathcal{R}^2$  due to curvature effects, thus violating the area law for sufficiently large values of the coupling constant $\xi$. The total entropy decreases for large field masses, with nonnegligible effects for $m > 1$ TeV, for our choice of parameters.}
    \label{figsds}
\end{figure*}

We notice that in Eq. \eqref{discham} there is no explicit dependence on the $m$ index, so we can rewrite the total Hamiltonian of the system as
\be \label{haml}
H= \sum_{l=0}^\infty (2l+1) H_l,
\ee
where $H_l$ is now in the form of Eq. \eqref{hoham}, up to the overall factor $a^{-1}$. Accordingly, the scalar field behaves as a system of coupled harmonic oscillators, where the coupling matrix $K$ is now given by 
\begin{align} \label{kmat}
K_{ij}=  & \left( \frac{l(l+1)}{i^2}+(m^2+\xi R) a^2  \right) \delta_{ij}  \notag \\
&+\frac{1}{i^2} \bigg[ \frac{9}{4} \delta_{i1} \delta_{j1} +\left(N-\frac{1}{2} \right)^2 \delta_{iN} \delta_{jN} \notag \\
&\ \ \ \ \ \ \ \ \ +\left( \left(i+\frac{1}{2} \right)^2+ \left(i-\frac{1}{2}  \right)^2   \right) \delta_{ij\ (i \neq 1,N)} \bigg] \notag \\
& -\left[\frac{(j+1/2)^2}{j(j+1)} \right] \delta_{i,j+1}  -\left[ \frac{(i+1/2)^2}{i(i+1)} \right] \delta_{i+1,j}. \notag \\
\end{align}
The last two terms denote nearest-neighbor interactions, originating from spatial derivatives of the field. 

Since we now have all the ingredients, inferred from our discretization procedure, we can turn to the computation of the scalar field entanglement entropy.


\section{Violating the area law from nonminimal coupling}   \label{sez4}

The ground state corresponding to the Hamiltonian \eqref{discham} is the direct product of the ground states of each $H_l$. Accordingly, the total entropy can be expressed in the form
\be \label{entscal}
\mathcal{S}(n,N)= \sum_l (2l+1) \mathcal{S}_l(n,N).
\ee
Here $n$ denotes the number of degrees of freedom which are traced out, i.e., those inside a sphere of radius $na$ centered at the origin.

We notice that, for $l \gg N$, the nondiagonal elements of the $K$ matrix in Eq. \eqref{kmat} are much smaller than its diagonal ones: one can set up a perturbative treatment to show that Eq. \eqref{entscal} converges, thus ensuring a finite entanglement entropy\footnote{See \cite{rev2} for more details on the perturbative expansion of the $K$ matrix for large angular momenta.}.

We will set a momentum cutoff $l_{\rm max}$ that ensures negligible error for the total entropy \cite{inver}. Let us choose for the moment $\xi=0$, which represents the minimal coupling case. Defining the radius $\mathcal{R}=a(n+1/2)$, we can see from Fig. \ref{fnocurv} that the scalar field entanglement entropy is proportional to $\mathcal{R}^2$, thus satisfying an area law. As already noted \cite{rev2, inver}, larger scalar field masses result in smaller entanglement amount, thus changing the proportionality coefficient between entanglement entropy and area. The presence of field-curvature coupling can alter this scenario, as we  now show in detail.


\subsection{Schwarzschild-de Sitter spacetime} \label{sez4A}

The Schwarzschild-de Sitter metric is the simplest solution in general relativity containing both a black hole event horizon and a cosmological event horizon \cite{SDS}. Accordingly, it provides an interesting background to study the entropy of black holes embedded in an expanding universe \cite{gibb, houche, jaco, asht, math, good}.

\begin{figure*}
 \centering
\includegraphics[scale=0.63]{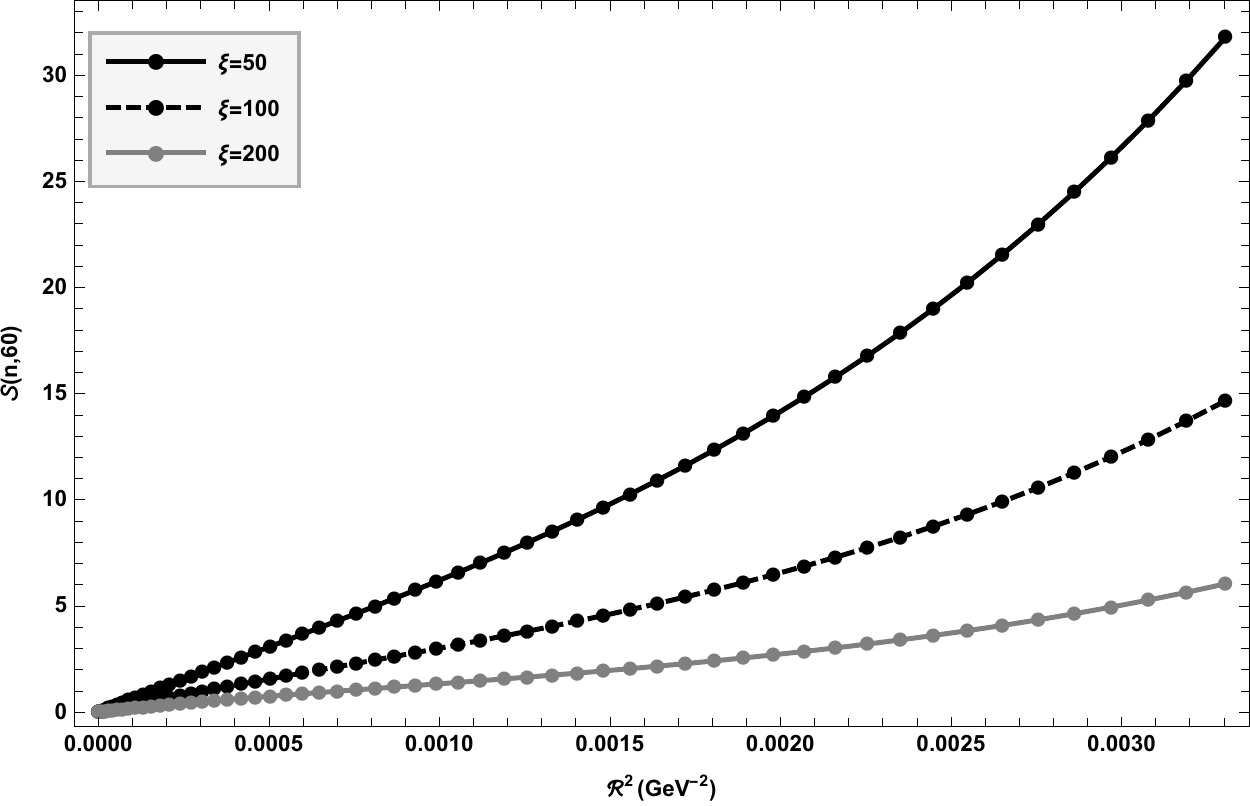}
\hspace{1cm}
\includegraphics[scale=0.63]{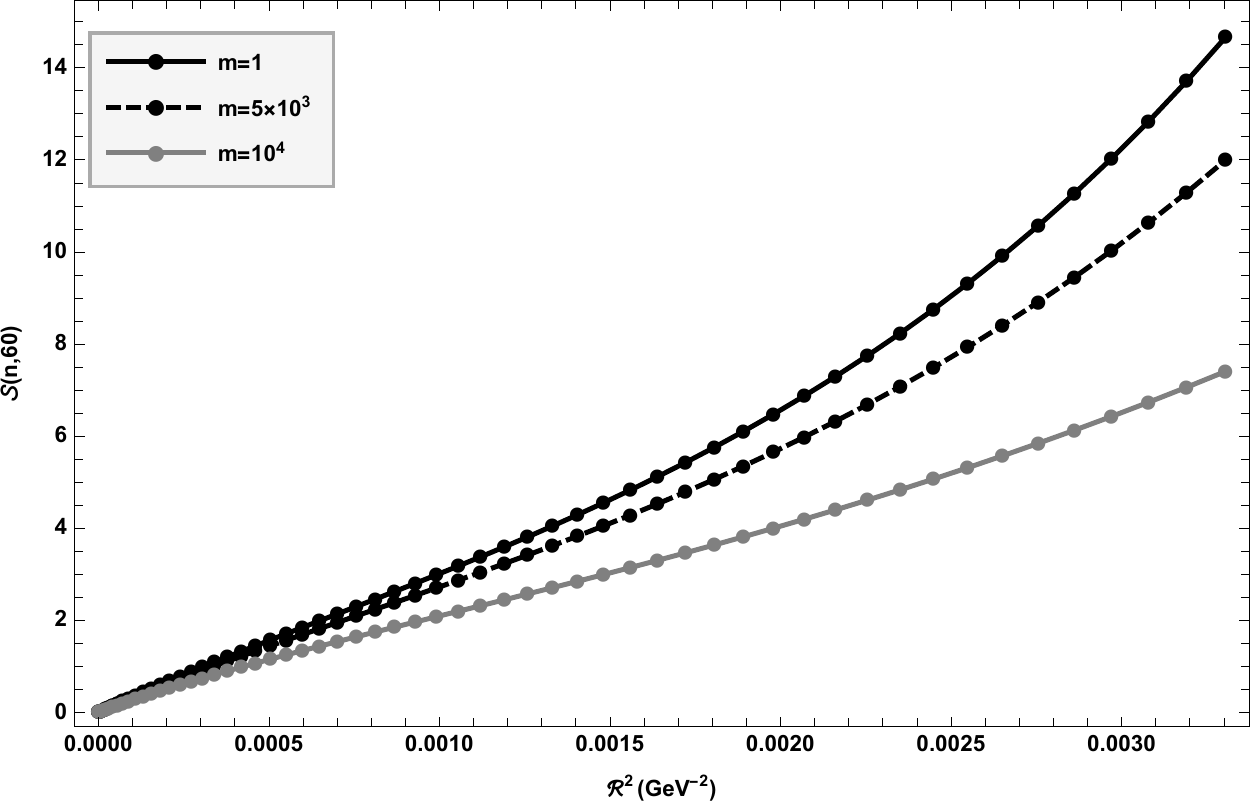}
    \caption{Entanglement entropy $\mathcal{S}(n,N)$ in Hayward spacetime for different values of the field-curvature coupling constant $\xi$ (on the left, with $m=1$) and field mass (on the right, with $\xi=10^2$). The other parameters are the same as in Fig. \ref{figsds}, with $b=2a$. Violation of area law is evident as $n \rightarrow N$, where the metric is less influenced by the de Sitter core.}
    \label{fighay}
\end{figure*}

In static coordinates, the Schwarzschild-de Sitter line element can be written as in Eq. \eqref{linel}, with
\be \label{sds}
f(r)= 1-\frac{2G M}{r}- \frac{r^2}{L^2},
\ee
where $M$ is the mass of the black hole and $L^2$ is proportional to the inverse of the positive cosmological constant driving cosmic expansion in the model. Assuming $3 \sqrt{3} GM/L < 1$, one finds two event horizons at \cite{good}
\be \label{bhor}
r_{\pm}= \frac{2 L}{\sqrt{3}} \cos \left[ \frac{1}{3} \cos^{-1} \left( \frac{3 \sqrt{3} G M}{L} \right)  \pm \frac{\pi}{3}  \right],
\ee
with $r_+ < r_-$. In Fig. \ref{figsds} we show the entanglement entropy for a nonminimally coupled scalar field in Schwarzschild-de Sitter spacetime, assuming $a N < r_+$, i.e, considering only scalar field degrees of freedom inside the inner horizon\footnote{We need to avoid $f(r)=0$, since in this case Eq. \eqref{lemt} diverges and thus we are not able to properly fix time. We also expect area law to be gradually restored at $r > r_+$, because in this limit $f(r)$ approaches de Sitter regime, as discussed below.}. We notice that close to the black hole singularity $r=0$ the entanglement entropy does not scale linearly with $\mathcal{R}^2$, due to curvature effects\footnote{We are neglecting the first shell, corresponding to $n=1$, in order to avoid divergences when computing discrete derivatives as in Eq. \eqref{fsec}.}. This implies a \emph{violation of the area law} behaviour, which is more evident at larger $\xi$, and suggests that local interactions between a quantum field and the underlying spacetime geometry can affect entanglement generation. Similar results are obtained at smaller $\xi$ if the parameter $L$ is also sufficiently small, i.e, for larger values of the cosmological constant, which can be typical of inflationary scenarios \cite{mae2, futa}. Slight deviations from area law are also found at $n \simeq N$: however, these simply represent boundary effects related to the discretization procedure, as discussed for example in \cite{qft2}. We also notice that for sufficiently large field masses the total entropy is significantly reduced, since in this case the mass term in the potential has a magnitude comparable to the coupling one.

\subsection{de Sitter spacetime}
In the limit $M=0$, we recover pure de Sitter spacetime, with constant scalar curvature $R=12/L^2$. In de Sitter background, field-curvature coupling simply adds a constant mass-like term, thus preserving the linear dependence of entanglement entropy on the boundary area (cfr. Fig. \ref{fnocurv}). As shown in Fig. \ref{figsds}, area law is also recovered for the Schwarzschild-de Sitter solution as $n \rightarrow N$, i.e., when the last term of Eq. \eqref{sds} becomes dominant and thus the scalar curvature approaches the constant de Sitter value.

\subsection{Hayward spacetime}

More recently, some modifications to the Schwarzschild-de Sitter solution have been proposed, with the aim of including a vacuum energy contribution into regular spacetime configurations that exhibit a stable behavior at small radii. To this end, the Hayward solution represents a widely investigated example of these models \cite{hay, hay2, hay3}, providing 
\begin{equation} \label{hayw}
f(r)=1-\frac{2GMr^2}{r^3+2GMb^2}.
\end{equation}
This ansatz yields the Schwarzschild solution at large distances, while fixing $b^2=1/\Lambda^2$ to guarantee that for $r\rightarrow0$ we find the de Sitter spacetime with cosmological constant $\Lambda$. Since $b$ should describe the approximate length below which quantum gravity effects dominate, we will assume it close to the ultraviolet cutoff, $b \simeq a$. At the same time, in order to have a regular nonextreme black hole, we require $M>M_*$, where 
\be \label{critm}
M_*= \frac{3 \sqrt{3} b}{4 G}
\ee
is the critical mass defined in \cite{hay}. In Fig. \ref{fighay} we show the scalar field entanglement entropy in Hayward spacetime, again changing the value of the coupling constant $\xi$ and the field mass $m$. In this case, we notice that deviations from area law are more evident as $n \rightarrow N$, since in this limit we are moving away from the de Sitter core of the metric and thus the scalar curvature changes more rapidly. Accordingly, we expect a similar violation of area law also at larger $N$, provided we are still far from the limit $r \rightarrow \infty$, where the metric approaches Schwarzschild regime, resulting in negligible scalar curvature.


\section{Conclusions and perspectives} \label{sez5}

In this work, we investigated the area law in presence of a nonminimally coupled Lagrangian, assuming a scalar field to be coupled with spacetime curvature. In this respect, we explored a Yukawa-like coupling as interacting potential. Specifically, we analyzed spacetime configurations that involve vacuum energy and we distinguished between spacetimes that are either singular, namely black holes, or regular, namely regular black holes. 

We motivated our choices, emphasizing that these geometric configurations hold great significance in cosmological scenarios such as inflation, reheating, etc. and black hole physics. We thus related our findings to the current comprehension of spherically-symmetric metrics that can be adapted to regimes of strong gravity. Indeed, we here considered the well-known Schwarzschild, Schwarzschild-de Sitter singular spacetimes and the regular Hayward metric, also discussing de Sitter space as a limiting case. Through our investigation, we demonstrated that nonminimal coupling can significantly influence the dependence of entropy on the area, suggesting that local interactions between quantum fields and the background geometry play a crucial role in entanglement generation. In particular, we showed that in Schwarzschild-de Sitter spacetime area law is violated close to the singularity, namely for $n \rightarrow 1$, while it is restored at larger radii. On the contrary, for Hayward spacetime we observed deviations as $n \rightarrow N$, when the radius of the region that is traced out is not too small. This suggests that departures from area law are more pronounced for regions where the scalar curvature exhibits fast variations as function of the radius. It is worth noting that we neglected possible back-reaction mechanisms at this stage, which could affect the dynamics of the quantum fields in certain cosmological contexts. 

To quantitatively analyze these modifications, we numerically computed the entropy by discretizing the Hamiltonian derived from our field scenario. In particular, we modeled the system as a collection of harmonic oscillators on a lattice of spherical shells. 

As a direct consequence of our procedure, we made critical comparisons with previous findings obtained in flat spacetime contexts, highlighting the main deviations from the area law resulting from the coupling between the field and curvature. We observed that these effects appear more pronounced for large coupling constants or in case of relevant contributions from vacuum energy, which are characteristic of early universe scenarios. Our outcomes are displayed and numerically analyzed, up to the well-known computational limitations related to the discretization procedure employed. We also underlined the physical consequences of our results and we showed that the influence of vacuum energy, in general, appears crucial in area law deviations, indicating that a relativistic field theory embedded in curved geometries might exhibit significant departures from flat spacetime results.

In future developments, we plan to further investigate additional couplings between fields and spacetime curvature, in order to better understand how interactions between fields and geometry affect area law. We also aim to analyze the role of spacetime symmetries and determine whether they can lead to violations of the area law. In this respect, possible generalizations of the above techniques to nonspherical entangling surfaces also requires further consideration. Finally, we plan to extend our study to different field theory contexts, including Dirac and/or tachyonic fields, K-essence models, Horndeski gravity, scalar-tensor theories, and more.

\section*{Acknowledgements}
OL acknowledges Grant No. AP19680128 from the Science Committee of the Ministry of Science and Higher Education of the Republic of Kazakhstan and he is also grateful to INAF, National Institute of Astrophysics, for the support and in particular to Roberto della Ceca, Gaetano Telesio and Filippo M. Zerbi for discussions. SM acknowledges financial support from “PNRR MUR project
PE0000023-NQSTI”.


\end{document}